\documentclass[reprint,superscriptaddress,amsmath,amssymb,aps,prb]{revtex4-1}
\usepackage{graphicx} 
\graphicspath{{Pictures/}}
\usepackage{hyperref}
\hypersetup{colorlinks = true, citecolor = blue, breaklinks = true}
\usepackage{xcolor}
\usepackage{multirow}
\usepackage[resetlabels]{multibib}
\newcites{SI}{References}
\usepackage[english]{babel}

\begin{document}

\title{Anion-order driven polar interfaces at LaTiO$_2$N surfaces}

\author{Silviya Ninova}
\affiliation{Department of Chemistry and Biochemistry, University of Bern, Freiestrasse 3, CH-3012 Bern, Switzerland}

\author{Ulrich Aschauer}
\email{ulrich.aschauer@dcb.unibe.ch}
\affiliation{Department of Chemistry and Biochemistry, University of Bern, Freiestrasse 3, CH-3012 Bern, Switzerland}

\begin{abstract}
	Perovskite oxynitrides have recently attracted attention for their ability to photocatalytically split water. Compared to oxides the arrangement of anions in the material represents a further structural degree of freedom. The bulk oxynitride LaTiO$_2$N prefers a bonding-dominated \textit{cis} nitrogen arrangement, while we have previously shown that the (001) surface prefers a non-polar \textit{trans} order to compensate polarity. Here we consider, using density functional theory calculations, the polar/non-polar interface that would necessarily be present between the two anion orders. We show that the Ti-terminated surface will adopt up to two \textit{trans} ordered surface layers, which has a beneficial effect on the oxygen evolution efficiency. We then consider the hypothetical case of a polar \textit{cis} ordered surface layer atop a non-polar \textit{trans} bulk and show that similar electronic reconstructions as in the LaAlO$_3$/SrTiO$_3$ interface can be expected when interfaces between different anion orders are engineered in one and the same oxynitride material.
\end{abstract}

\maketitle

%%%MAIN TEXT%%%%
%%%%%%%%%%%%%%%%%%%%%%%%%%%%%%%%%%%%%%%%%%%%%%%%%%%%%%%%%%%%%%%%%%%
\section*{Introduction}
%%%%%%%%%%%%%%%%%%%%%%%%%%%%%%%%%%%%%%%%%%%%%%%%%%%%%%%%%%%%%%%%%%%
Oxynitrides have recently attracted attention as electrodes for photocatalytic water-splitting under visible light.\cite{Ebbinghaus2009, Takata2015, Ahmed2016} These mixed-anion compounds are semiconductors and combine the characteristics of the corresponding pure oxides and nitrides. The former have a good chemical stability, but their large band gaps - typically in excess of 3\,eV - limit the absorption of solar light to the ultra-violet region. The substitution of oxygen by less electronegative nitrogen reduces the band gap, however at the expense of a reduction in stability.

Oxynitrides can occur in the perovskite crystal structure where common deviations from the perfect cubic cell, such as rotations and distortions of the cation coordination octahedra, can have an impact on the electronic properties of the material.\cite{Kubo2016} In addition, perovskite oxynitrides are characterised by the order of the oxygen and nitrogen anions. While for some materials, such as LaZrO$_2$N and LaTiO$_2$N,\cite{Clarke2002} a complete disorder was reported, both experiment and theory have suggested the preference for a short-range \textit{cis} order of nitrogen in 2-dimensional planes\cite{Yang2011,Porter2014} that optimises the coordination of the cation by the anionic ligands.\cite{Yang2011}

\begin{figure}[t!]
	\centering
	\includegraphics[width=0.9\columnwidth]{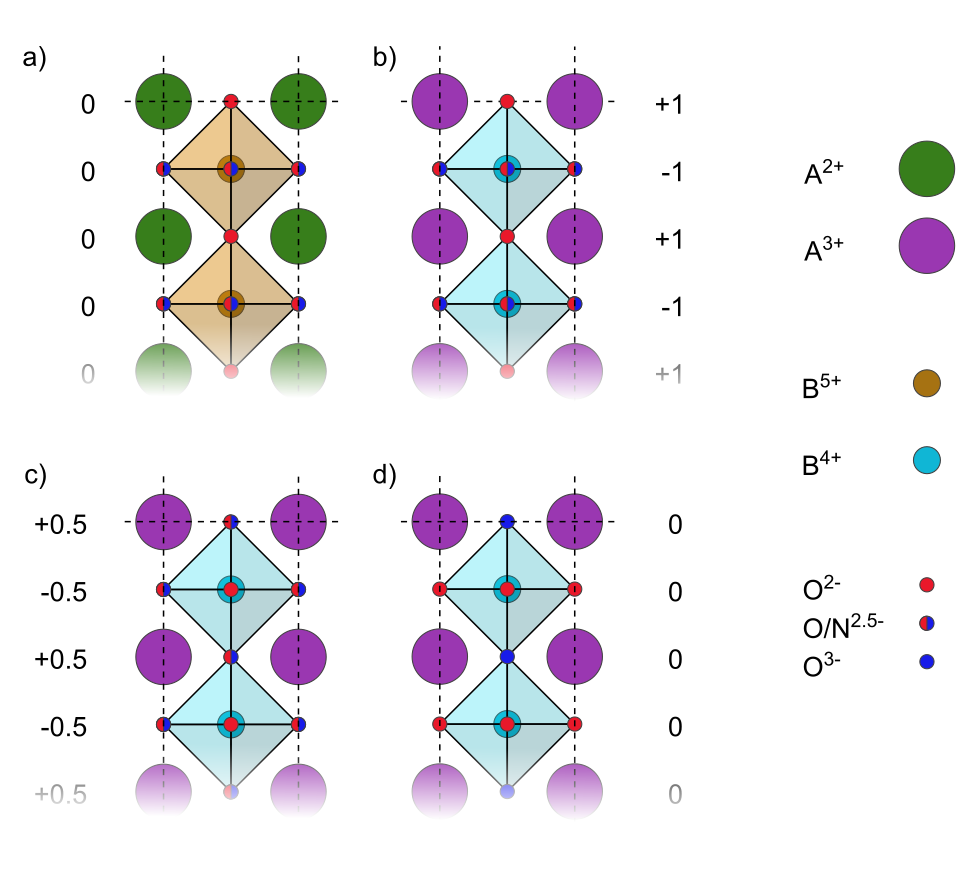}
	\caption{Schematic view of the electrostatics of different anion orders at oxynitride (001) surfaces: a) A$^{2+}$B$^{5+}$O$_2$N oxynitrides have charge neutral atomic (001) planes for a 2D in-plane \textit{cis} order that is also stable in the bulk. b) In a A$^{3+}$B$^{4+}$O$_2$N oxynitride the same in-plane \textit{cis} anion order leads to charged (001) planes. c) A partial reordering with N also in apical positions (\textit{cis} order along in- and out-of-plane direction) reduces the layer charge, while d) a complete \textit{trans} order results in charge-neutral planes.}
	\label{fig:schematic}
\end{figure}

This, however, may not necessarily be the case at the surface. Surfaces with alternating net positively and negatively charged layers are inherently unstable due to the diverging electrostatic energy and require electronic and/or structural reconstructions to cancel the resulting surface dipole. \cite{Goniakowski2007} Perovskite oxynitrides with the general formula A$^{2+}$B$^{5+}$O$_2$N can have a non-polar stacking of neutral A$^{2+}$O/B$^{5+}$ON planes along the {001} direction while preserving the bulk 2D \textit{cis} anion order in BON planes also at the surface (Fig. \ref{fig:schematic}a). Oxynitrides with the general formula A$^{3+}$B$^{4+}$O$_2$N, however, would result in alternating charged planes (A$_2^{2+}$ON/B$_2^{5+}$O$_3$N or A$_2^{2+}$O$_2$/B$_2^{5+}$O$_2$N$_2$ depending on the specific {001} direction with respect to the anion order, see Figs. \ref{fig:schematic} b and c), while preserving a \textit{cis} anion order. As we have previously shown for LaTiO$_2$N using density functional theory (DFT) calculations, a local \textit{trans} anion reordering at the surface results in a sequence of neutral A$^{3+}$N/B$^{4+}$O$_2$ layers (Fig. \ref{fig:schematic} d) and lower surface energies compared to \textit{cis} ordered surfaces.\cite{Ninova2017} 

This implies that at surfaces of A$^{3+}$B$^{4+}$O$_2$N oxynitrides such as LaTiO$_2$N, the electrostatic energy will favour a \textit{trans} nitrogen order along the out-of-plane direction over the bonding-dominated stable \textit{cis} order of the bulk. This change in anion order at the surface will result in a \textit{trans} region at the surface and an underlying \textit{cis} bulk. In the present paper, we investigate the structure and stability of this \textit{cis}/\textit{trans} interface, the thickness of the \textit{trans} layer and the resulting surface properties of LaTiO$_2$N (001) using DFT calculations. As the interface resembles the LaAlO$_3$/SrTiO$_3$ interface that hosts a 2D electron gas to compensate polarity,\cite{Ohtomo2004} we also investigate the hypothetical case of a \textit{trans} bulk with a \textit{cis} surface to assess if an anion-order induced non-polar/polar interface could lead to similar effects in an oxynitride material.

%%%%%%%%%%%%%%%%%%%%%%%%%%%%%%%%%%%%%%%%%%%%%%%%%%%%%%%%%%%%%%%%%%%
\section*{Methods}
%%%%%%%%%%%%%%%%%%%%%%%%%%%%%%%%%%%%%%%%%%%%%%%%%%%%%%%%%%%%%%%%%%%

Initial LaTiO$_2$N (001) surface slabs were built from two different bulk models, corresponding to model 3 and model 4 in Ref. \cite{Ninova2017}. Both bulk models have a \textit{cis} arrangement of Ti-N bonds, which is the most stable anion order in the bulk, but differ in the direction of the Ti-N chains, which propagate in \textit{bc} (model 3) and \textit{ab} (model 4) planes respectively. The resulting slabs thus differ in their atomic-layer chemical composition - either La$_2$NO/Ti$_2$O$_3$N or La$_2$O$_2$/Ti$_2$O$_2$N$_2$ - and thus in their resulting surface dipole moment. From now on, we will refer to them as model A and model B respectively. 

Our surface slabs are asymmetric, 2$\times$2 perovskite unit cells wide and 4 unit-cell (8 atomic layers) layers thick. Periodic images along the surface normal are separated by 10\,\r{A} of vacuum and a dipole correction\cite{Bengtsson1999} is applied, so as to cancel the effect of the artificial electric field across the slab as a result of the differently charged terminations. The atoms in the bottommost unit cell are kept fixed at their bulk position and relaxations are performed with convergence thresholds of $1.4\cdot10^{-5}$\,eV for the energy and 0.05\,eV/\r{A} for the forces.

\begin{figure}
	\centering
	\includegraphics[width=0.9\columnwidth]{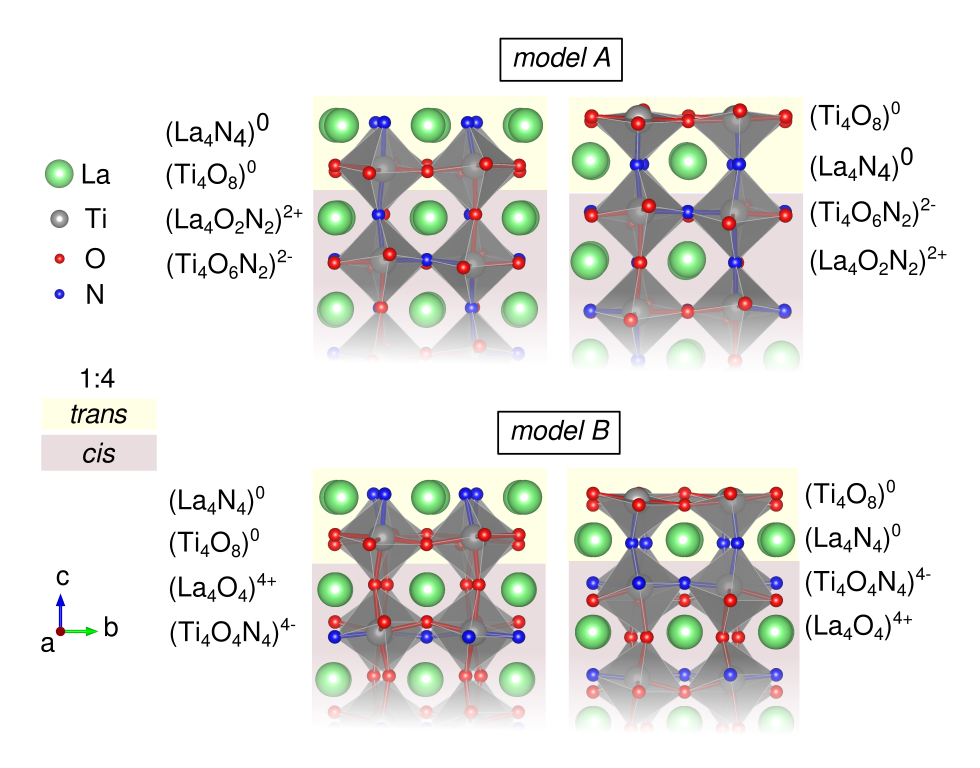}
	\caption{1:4 surface slab models with the topmost unit-cell layer having reordered \textit{trans} composition while the underlying layers are \textit{cis} ordered.}
	\label{fig:models}
\end{figure}

Based on these initial model A and model B slabs, we construct mixed anion-ordered slabs by exchanging N and O atoms within one unit-cell layer at a time, starting from the topmost one and proceeding downwards, resulting in the electroneutral La$_2$N$_2$/Ti$_2$O$_4$ stacking sequence in the anion-exchanged layers. Notation-wise, such slabs will be called ``1:4'' when the first layer of the slab is \textit{trans}-ordered, ``2:4'' - when the top half of the slab trans-ordered and ``3:4'' - when only the bottommost unit-cell is \textit{cis}-ordered. Following this notation for the number of layers with altered (\textit{trans}) anion order compared to the total thickness of the slab, the initial purely \textit{cis}-ordered surface slabs are ``0:4''. Examples of model A and model B ``1:4'' slabs with both La and Ti termination layers are shown in Fig. \ref{fig:models}.

The energy gain/loss of this anion exchange, E$_{ionex}$, is evaluated as the energy difference between relaxed surface slabs with mixed anion order and without. Additionally, we assess the quantitative effect of the anion exchange on surface energies, following equation \ref{eq:surf-en}. We assume that the process of surface creation with anion reordering happens in two steps - first, the surface is cleaved from the bulk structure and then the anion exchange occurs, which is the reason why $E^{urlx}$ is chosen to be that of models A or B with no anion exchange; subsequently, the surface relaxes into a slab with the chosen anion order. E$_{bulk}$ is the bulk energy of each model and A$_{surf}$ is the surface area.

\begin{equation}
E_{surf}^{top}=\frac{2E_{slab}^{rlx}-E_{slab}^{urlx}-E_{bulk}}{2A_{surf}}\label{eq:surf-en}
\end{equation}

We carry out all calculations with the PBE functional\cite{PBE} and the Quantum ESPRESSO package.\cite{QE-2009, QE-2017} The ion-electron interactions are described with ultrasoft pseudopotentials,\cite{UltraSoftPseudo} and the wavefunctions are expanded in plane waves. The chosen cut-offs for the kinetic energy and the augmented density are 40\,Ry and 320\,Ry, respectively. A Hubbard U parameter of 3\,eV is applied on the Ti-3\textit{d} orbitals.\cite{Anisimov1991} The Brillouin zone is sampled with a sufficiently dense 6$\times$6$\times$1 Monkhorst-Pack k-mesh.\cite{Monkhorst1976} The projected density of states (PDOS) are plotted with a Gaussian broadening of 0.01\,Ry. Finally, we use VESTA for all graphical representations of our models.\cite{Momma2011}

%%%%%%%%%%%%%%%%%%%%%%%%%%%%%%%%%%%%%%%%%%%%%%%%%%%%%%%%%%%%%%%%%%%
\section*{Results and Discussion}
%%%%%%%%%%%%%%%%%%%%%%%%%%%%%%%%%%%%%%%%%%%%%%%%%%%%%%%%%%%%%%%%%%%
%%%%%%%%%%%%%%%%%%%%%%%%%%%%%%%%%%%%%%%%%%%%%%%%%%%%%%%%%%%%%%%%%%%
\subsection*{Mixed-order surfaces}
%%%%%%%%%%%%%%%%%%%%%%%%%%%%%%%%%%%%%%%%%%%%%%%%%%%%%%%%%%%%%%%%%%%

\begin{figure}
  \centering
  \includegraphics[width=0.9\columnwidth]{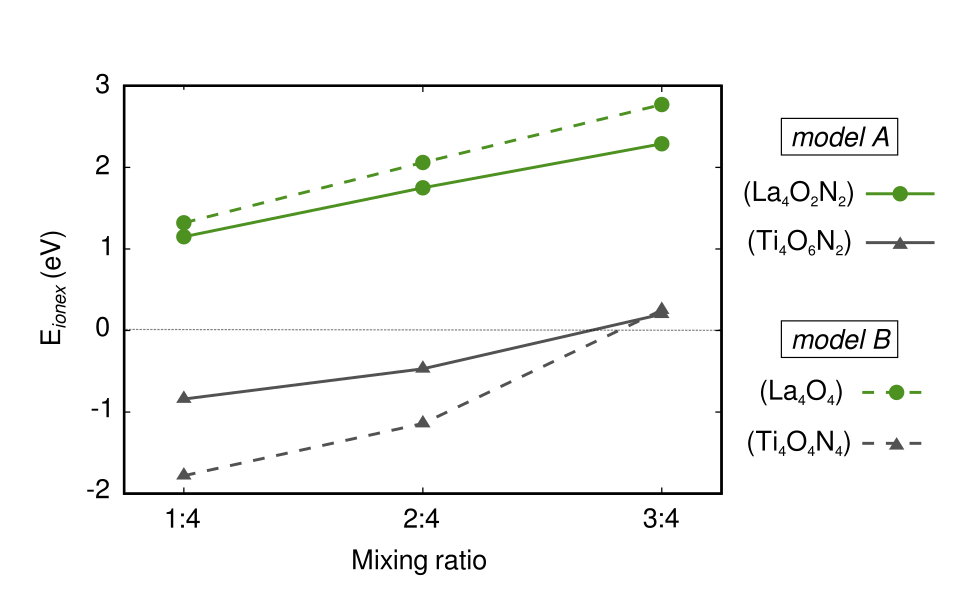}
  \caption{The anion exchange energies (E$_{ionex}$), calculated as the difference in total energy between the relaxed models without and with anion exchange.}
\label{fig:exch-en}
\end{figure}

In figure \ref{fig:exch-en}, we report the energy gain (negative) or loss (positive) associated with exchanging anions in successive layers for both the La and the Ti terminated surfaces of models A and B. We observe that assuming a \textit{trans} order has dramatically different effects for the La and Ti terminations, independently of the model. While the \textit{trans} (TiO$_2$)/(LaN) surface order is favourable for Ti-terminated surfaces (negative E$_{ionex}$), no such stabilization occurs for the La-terminated surface. This energy evolution is also reflected in the surface energies as shown in Table \ref{tbl:surf-en-2x2x4}. For the Ti terminated surface, we observe negative E$_{ionex}$ for anion exchange in the first and second layer, reordering being more favourable for the more polar model B. An anion exchange from the 3rd layer onwards is energetically no longer favorable, suggesting a \textit{trans} layer thickness of 2 unit cells at the Ti terminated surface. For the La terminated surface, on the other hand, we observe positive and continuously increasing E$_{ionex}$, implying that the \textit{cis} ordered surface is most stable. This is surprising, considering that the surface slab with the lowest surface energy was found to have a LaN termination,\cite{Ninova2017} the difference being that there the whole slab assumes a \textit{trans} anion order.

We perform extensive tests to ensure that these observations are not affected by the computational setup. Indeed, neither the thickness nor the type of slab (symmetric or asymmetric) alters the observed trends (see Tables \ref{tbl:exch-en-all} and \ref{tbl:surf-en6}). The frozen bottommost layer does also not significantly alter the slab geometry, as shown in Figure \ref{fig:displacements-sq}. We thus assume that the observed energy evolutions and deviating behaviour for the terminations are physically meaningful. As anion exchange affects the structure and energetics in more than one way, we will, in following paragraphs, try to decouple these energy contributions and investigate the possible cause for the different behaviour of the two terminations.

The most obvious effect is that here we compute a \textit{trans} surface layer on top of a \textit{cis} ordered bulk and thus impose the \textit{cis} lattice parameters on the \textit{trans} layer. Due to different Ti-N and Ti-O bond lengths, this will lead to strain in the \textit{trans} layer that could affect the energy. By relaxing the in-plane lattice parameters of a model B 1:4 slab, we  find a 0.2\% smaller cell parameter that, however, amounts to only 0.02\,eV strain energy per exchanged layer. Also this strain effect should be similar for both terminations and can therefore not explain the observed marked difference.

Relaxation of the internal geometry could also contribute to the stabilization observed for Ti-terminated surfaces. While geometry changes for the Ti terminated surfaces are generally small, the model B La terminated 1:4 surface undergoes an expansion of the topmost layer by 4.6\% with respect to 0:4 (see Table \ref{tbl:surf-geo}) without large changes in the underlying layers. We also observe a change in the octahedral rotation pattern at the La terminated surface, where the longer Ti-O bonds lead to out-of-phase rotations along the surface normal, while the aforementioned expansion of the surface reduces the tilt angle from 12.4 to 9.11$^{\circ}$. The Ti terminated surface retains the bulk rotation pattern and has smaller changes in tilt angle from 13.0 to 11.9$^{\circ}$. Geometry changes for the Ti terminated surface are thus small, yet we find negative reordering energies for the Ti terminated slabs of both models only after geometry optimization (see Table \ref{tbl:surf-en-2x2x4}). This is also observed for the 1:4 surface of the less polar Model A. More interesting though is the 1:4 surface in Model B, where anion exchange in the topmost unit cell even without geometry relaxation results in a stabilization by 0.12\,eV compared to the purely \textit{cis}-ordered surface. While geometry relaxations thus play a role, the last observation points to a strong alteration of the electrostatics in the slab.

\begin{figure}[t]
  \centering
  \includegraphics[width=0.9\columnwidth]{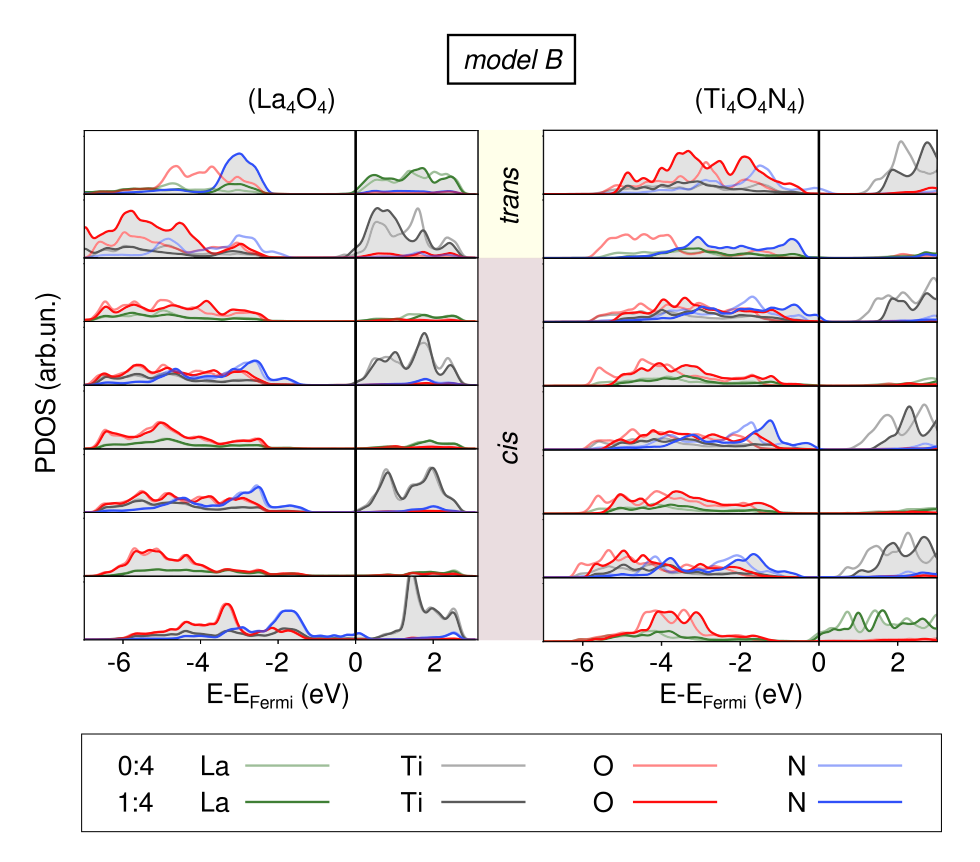}
  \caption{Layer-resolved PDOS of both terminations in model B, where results for the 1:4 slab (filled in gray) are compared with those for the 0:4 slab (unfilled curves). The PDOS for Model A are shown in Figure \ref{fig:pdos-A}.}
\label{fig:pdos-B}
\end{figure}

Anion exchange leads not only to the formation of polar-compensated atomic layers at the surface, which should be favoured since they reduce the surface dipole, but also introduces a polar/non-polar interface at the \textit{trans}/\textit{cis} boundary. Strong electronic reconstructions, such as the accumulation or depletion of charge, are expected to arise in this region, affecting the electrostatic interactions in the slab. To establish these electronic structure changes, we examine the layer-resolved projected density of states (PDOS) of the 0:4 and 1:4 slabs for both terminations of model B shown in Figure \ref{fig:pdos-B} and model A in Figure \ref{fig:pdos-A}. For the pure \textit{cis} 0:4 slab N-2\textit{p} states constitute the top of the valence region, while the bottom of the conduction band is dominated by Ti-3\textit{d} states. As a result of the nominally charged atomic layers, an electric field appears through the slab, the stabilisation of which requires electronic reconstruction via a compensating charge. In the case of the 0:4 (La$_4$O$_4$)-termination this occurs by partially filling the Ti-bands in the first sub-surface layer. The 0:4 (Ti$_4$O$_4$N$_4$)-termination, on the other hand, is hole doped in the surface layer. This can be observed more clearly for the case of the more polar model B, on which we will focus our discussion in the following.

In the La-terminated slab, anion exchange from 0:4 to 1:4 leads to significant PDOS changes only in the topmost three atomic layers. We observe bonding-induced changes in the nitrogen and oxygen states of the valence band: Due to the lack of strong bonds with Ti in the 1:4 structure, the N-$p_x$ and N-$p_y$ states form a large peak at about -3\,eV and the valence band maximum is no longer N-2\textit{p} dominated. Despite these changes, there is, however, no significant impact on the electron doping of the subsurface (Ti$_4$O$_4$)$^{0}$ layer, which does not disappear or shift to lower layers with the formation of the charge-neutral surface layers upon anion exchange. In contrast, anion exchange in the Ti-terminated surface affects the whole slab down to the bulk layers: Besides a notable upward shift of states in all atomic layers, the hole doping shifts from the surface layer to the \textit{trans}/\textit{cis} interface. By assuming that charge accumulation layers can be represented by charged plates, we can make an analogy to a capacitor model, where the energy decreases with decreasing plate distance. The absence of a shift in the charge accumulation layers for the La-terminated slab implies no marked change in energy, while the shift of one unit-cell layer for the Ti-terminated surface results in a marked reduction in energy. It is thus only for the Ti-terminated surface that the electrostatic energy gain can outweigh the other energy contributions, resulting in an overall stabilization. It is necessary, however, to point out that such a capacitor analogy, while valid for our slabs, may lose its validity when macroscopic dielectric materials are considered.

\begin{figure}
  \centering
  \includegraphics[width=0.9\columnwidth]{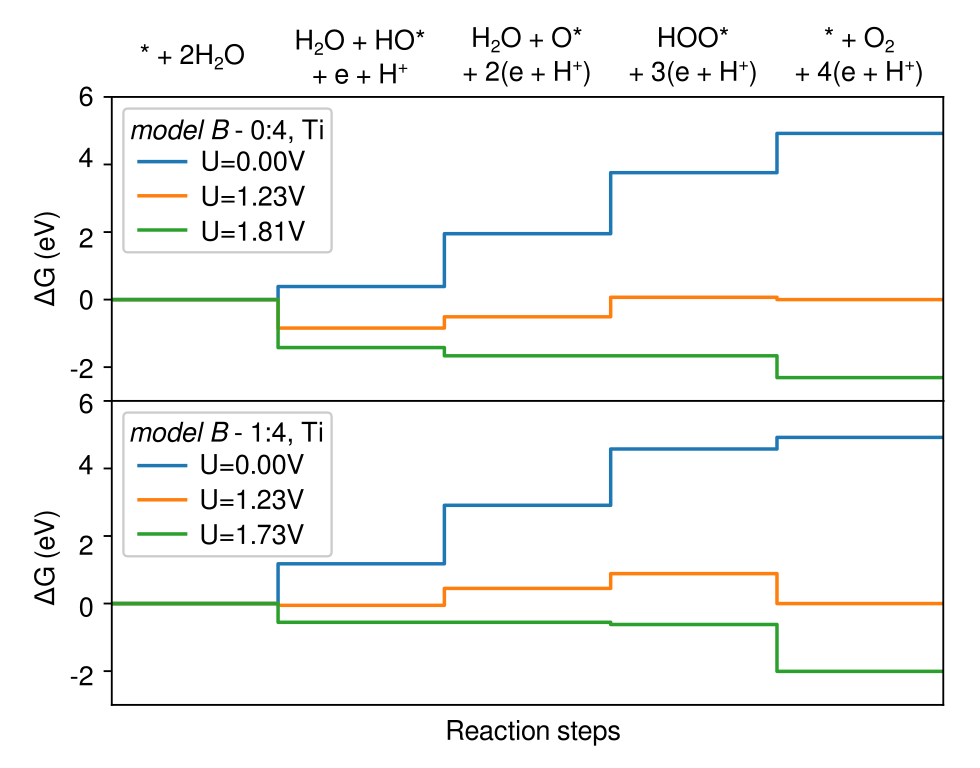}
  \caption{Free energy profile of the oxygen evolution reaction on the purely \textit{cis} (0:4, top) and anion-exchanged (1:4, bottom) Ti-terminated model B. The reaction energies were reported in absence of an applied bias potential, at equilibrium 1.23\,V and at the lowest potential, which renders all steps exothermic.}
\label{fig:norskov}
\end{figure}

Polar/non-polar interfaces with accompanying internal electric fields were postulated to be promising for photocatalysis due to the effective exciton dissociation and spatial separation of anode and cathode reactions. \cite{Guo2016} We, therefore, investigate the influence of such a non-polar/polar interface on the water-oxidation efficiency. In this reaction molecular oxygen is formed via four proton-coupled electron transfer steps. To compare the efficiency of the 0:4 and 1:4 surface, we compute the reaction free energies of the four steps following the procedure proposed by N{\o}rskov and co-workers\cite{Norskov2004,Valdes2008} and determine the overpotential necessary to render the reaction thermodynamically favorable (see Figure \ref{fig:norskov}). We observe a slight decrease in overpotential from 0.58\,eV for 0:4 to 0.50\,eV in the case of 1:4. The different electric field in the 0:4 and 1:4 do not affect the interaction with the intermediates (OH*, O*, OOH*) in terms of bond lengths with the Ti adsorption site. Interestingly though, the potential determining step changes from the formation of the hydroperoxo adsorbate for 0:4 to the deprotonation of the hydroxo for 1:4. Based on these results, we expect the Ti-termination with mixed anion order to be more reactive for water splitting than a pure \textit{cis} termination.

%%%%%%%%%%%%%%%%%%%%%%%%%%%%%%%%%%%%%%%%%%%%%%%%%%%%%%%%%%%%%%%%%%%
\subsection*{The polar interface}
%%%%%%%%%%%%%%%%%%%%%%%%%%%%%%%%%%%%%%%%%%%%%%%%%%%%%%%%%%%%%%%%%%%

We have in the previous section established that at the Ti-terminated surface of LaTiO$_2$N, anion reordering leads to a non-polar \textit{trans} surface stacking atop a polar \textit{cis} bulk anion order. The interface between the two regions bears resemblance with polar/non-polar interfaces obtained by coherent epitaxially growth of polar materials atop non-polar substrates as for example in the LaAlO$_3$/SrTiO$_3$ (LAO/STO) interface. \cite{Ohtomo2002,Biscaras2010} In these structures the interface leads to a diverging electrostatic energy (the ``polar catastrophe''), which is stabilized by charge transfers from the surface of the polar overlayer to the interface, forming a so-called 2-dimensional electron gas (2DEG). \cite{Ohtomo2004,Goniakowski2007,Goniakowski2014} The interfaces of type TiO$_2^{0}$/LaO$^{+1}$ and SrO$^{0}$/AlO$_2^{-1}$ give rise to electron (n-type) and hole (p-type) accumulation respectively at the interface once the polar overlayer has reached a critical thickness, which is 4 and 6 unit-cells for the n-type and p-type interface respectively.\cite{Ohtomo2004,Guo2016,Wang2018}
 
There are two differences between the LaTiO$_2$N case discussed above and the LAO/STO interface: first, the reordered surface layer on LaTiO$_2$N (001) is non-polar, whereas the bulk underlayer is polar and second, the interface considered here is not between two different materials, but within the same material, differing only by its anion order. To assess the potential of using anion order to construct interfaces similar to LAO/STO, we consider in the following section the hypothetical scenario of a \textit{trans}-bulk/\textit{cis}-surface. While such a structure is not expected to arise from surface stabilization, our findings motivate this computational experiment, in particular since engineering such anion-ordered layers does not seem impossible using modern thin-film growth techniques. Apart from direct growth of the different layers from nitride and oxide targets, it was shown for example that compressive epitaxial strain can lead to \textit{trans} order in SrTaO$_2$N and LaTiO$_2$N.\cite{Oka2014,Oka2017,Vonrueti2018}. Once a film becomes thick enough for the effect of strain to weaken, a \textit{cis} ordered overlayer would result, leading to the thought-after interface.

In our case, the n-type interface should result at the boundary between (La$_4$O$_4$)$^{4+}$/(Ti$_4$O$_8$)$^{0}$, while the p-type interface is formed at (Ti$_{4}$O$_{4}$N$_{4}$)$^{4-}$/(La$_{4}$N$_{4}$)$^{0}$ boundaries. The thickness of formally electroneutral \textit{trans} layers are kept fixed to 4 unit cells, whereas we test an increasing thickness for the \textit{cis}-ordered part of the slab, from 2 to 6 unit cells, so as to examine the point of band inversion. Due to the increased number of atoms, we use a ($\sqrt{2}/2 \times \sqrt{2}/2$)-R45$^{\circ}$ cell with respect to our above calculations. No atoms are frozen and we used cell dimensions of the \textit{trans} anion arrangement (5.539$\times$5.608\,\AA \,in the \textit{xy}-plane) and a constant vacuum of 12 \AA\, along the surface normal. As mentioned above, we find the strain effect resulting from lattice mismatch between the two regions to be negligible.

\begin{figure}[h]
  \centering
  \includegraphics[width=0.9\columnwidth]{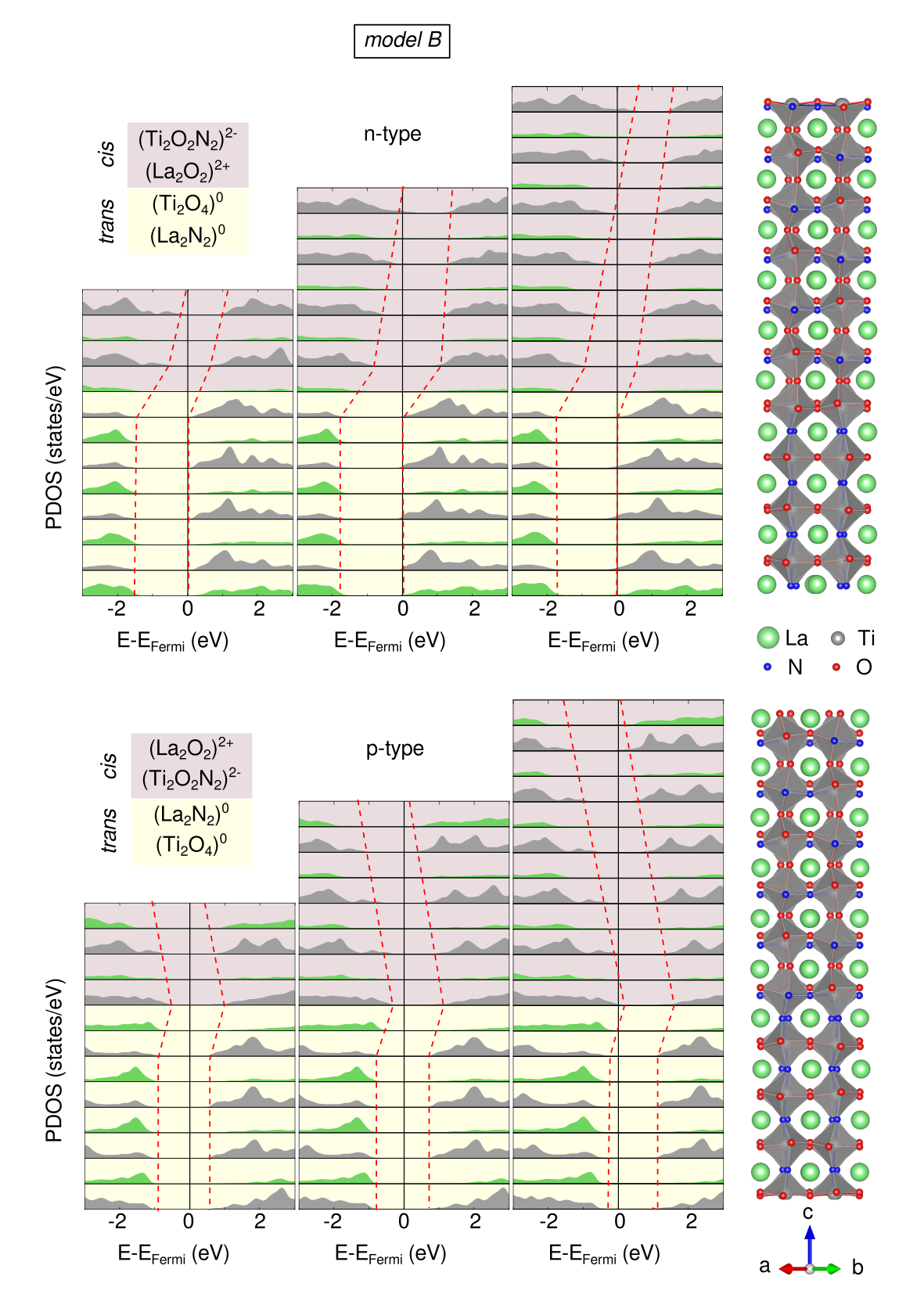}
  \caption{Layer-resolved PDOS of model B interfaces with different thickness of the \textit{cis}-ordered surface layers. The dashed red lines mark the band edges.}
\label{fig:pdos-B-laosto}
\end{figure}

In Figure \ref{fig:pdos-B-laosto} we observe no charge transfers from the surface to the interface layers for a two unit-cell \textit{cis} overlayer. In agreement with the LAO/STO interface already for four and even more so for six \textit{cis} unit cells, we see hole doping of the surface and electron doping in Ti states in particular in layers below the interface. While the exact thickness of the 2DEG in n-type LAO/STO interfaces is still debated, there have been clear indications of it being hosted - at least partially - on interface Ti atoms.\cite{Pentcheva2006,Sing2009} In the present case, the 2DEG seems to be excluded from the very interface, which is possibly explained by changes in orbital energies due to polar distortions and octahedral rotations as outlined in the supporting information section \ref{sec:polar-interface}. Conversely, in the p-type interface (Figure \ref{fig:pdos-B-laosto}), we do not yet observe charge transfers for six \textit{cis} unit cells. However for the largest thickness investigated here, the surface conduction band is already very close to the Fermi level, suggesting imminent charge transfers for further increased thickness of the \textit{cis} layer. As shown in Figure \ref{fig:pdos-A-laosto}, interfaces constructed from the less polar \textit{cis} Model A do not reach the charge transfer state for the \textit{cis} layer thickness studied here. This could be ascribed to the lower charge per atomic layer and we would expect it to need at least double the amount of \textit{cis} unit cells compared to the more polar model B.

%%%%%%%%%%%%%%%%%%%%%%%%%%%%%%%%%%%%%%%%%%%%%%%%%%%%%%%%%%%%%%%%%%%
\section*{Conclusions}
%%%%%%%%%%%%%%%%%%%%%%%%%%%%%%%%%%%%%%%%%%%%%%%%%%%%%%%%%%%%%%%%%%%
In the present study we investigate the formation of a polar/non-polar interface at the LaTiO$_2$N (001) surface as a mechanism for surface stabilization. While the preferred bulk order is \textit{cis}, the Ti-terminated surface is predicted to be stabilized by up to two non-polar \textit{trans} layers. This effect is absent on the La-terminated surface, which we explain with different electronic reconstructions taking place in the two cases. We show that the overpotential for the oxygen evolution reaction is smaller on surfaces with this polar/non-polar interface compared to purely \textit{cis} ordered surfaces. Finally we have shown that anion order with a \textit{trans} bulk and a \textit{cis}-ordered surface can lead to electronic reconstruction effects and electron accumulation at the interface similar to the LaAlO$_3$/SrTiO$_3$ interface.

%%%%%%%%%%%%%%%%%%%%%%%%%%%%%%%%%%%%%%%%%%%%%%%%%%%%%%%%%%%%%%%%%%%
\section*{Acknowledgements}
%%%%%%%%%%%%%%%%%%%%%%%%%%%%%%%%%%%%%%%%%%%%%%%%%%%%%%%%%%%%%%%%%%%
We thank Claudine Noguera for helpful discussions. This research was funded by the SNF Professorship Grant $PP00P2\_157615$. Calculations were performed on UBELIX (http://www.id.unibe.ch/hpc), the HPC cluster at the University of Bern.

%%%END OF MAIN TEXT%%%


%\balance


%%%REFERENCES%%%
\bibliography{library.bib}

%%%%%%%%%%%%%%%%%%%%%%%%%%%%%%%%%%%%%%%%%%%%%%%%%%%%%%%%%%%%%%%%%%%
%%%                        SUPPLEMENTARY                        %%%
%%%%%%%%%%%%%%%%%%%%%%%%%%%%%%%%%%%%%%%%%%%%%%%%%%%%%%%%%%%%%%%%%%%

%reset all style and numbering
\clearpage
\clearpage
\setcounter{page}{1}
\renewcommand{\thetable}{S\arabic{table}}  
\setcounter{table}{0}
\renewcommand{\thefigure}{S\arabic{figure}}
\setcounter{figure}{0}
\renewcommand{\thesection}{S\arabic{section}}
\setcounter{section}{0}
\renewcommand{\theequation}{S\arabic{equation}}
\setcounter{equation}{0}
\onecolumngrid

%create title
\begin{center}
\textbf{\large Supplementary information for\\\vspace{0.5 cm}
\LARGE Anion-order driven polar interfaces at LaTiO$_2$N surfaces\\\vspace{0.3 cm}
\large by \\\vspace{0.3cm}
Silviya Ninova and Ulrich Aschauer}
\end{center}

%%%%%%%%%%%%%%%%%%%%%%%%%%%%%%%%%%%%%%%%%%%%%%%%%%%%%%%%%%%%%%%%%%%
\section{Surfaces}
%%%%%%%%%%%%%%%%%%%%%%%%%%%%%%%%%%%%%%%%%%%%%%%%%%%%%%%%%%%%%%%%%%%

In Table \ref{tbl:surf-en-2x2x4}, we show the surface and anion exchange energies for La and Ti terminations of surfaces constructed from both model A and B as a function of the number of exchanged layers. While for the La-terminated surface, there is a continuous increase of the surface energy, independent of the model, the surface energy for the Ti-termination drops below that of the purely \textit{cis} surface (0:4) for up to two exchanged layers (2:4). An analysis of the anion exchange energy without relaxation (E$_{ionex}^{urlx}$, Eq. \ref{eq:ionex_urlx}) shows that only the Ti-terminated surface of model B favourably exchanges one layer of anions, while relaxation effects (E$_{ionex}$) lead to the aforementioned preference for up to two exchanged layers for both models A and B.

\begin{eqnarray}
E_{ionex}^{urlx} &=& E^{urlx}(x:4) - E^{rlx}(0:4), \textrm{for x=(1,3)} \label{eq:ionex_urlx}\\
E_{ionex}^{rlx}  &=& E^{rlx}(x:4)  - E^{rlx}(0:4), \textrm{for x=(1,3)} \label{eq:ionex_rlx}
\end{eqnarray}

\begin{table}[h]
\caption{LaTiO$_2$N (001) surface energy ($E_{surf}$) and anion exchange energy ($E_{ionex}$) of both La and Ti terminations in the model slabs with different \textit{trans}:total ratios constructed from models A and B (see Figure \ref{fig:models}).}
\label{tbl:surf-en-2x2x4}
\centering
\begin{tabular}{c|c|cc|cc|cc}
\hline
\multirow{2}{*}{Termination}& \textit{trans}:total & \multicolumn{2}{c|}{$E_{surf}$\,(J/m$^2$)} & \multicolumn{2}{c|}{$E_{ionex}$\,(eV)} & \multicolumn{2}{c}{$E_{ionex}^{urlx}$\,(eV)}\\
\multirow{2}{*}{} 	& (layers) 	& model A	& model B	& model A	& model B & model A	& model B\\
\hline
\multirow{4}{*}{La}	& 0:4 &	1.161	&	1.255	& - 	& -		  & -		& -		 \\
\multirow{4}{*}{}	& 1:4 &	1.456	&	1.590	& 1.15 	& 1.32 	  & 2.25	& 5.40	 \\
\multirow{4}{*}{}	& 2:4 &	1.612	&	1.776	& 1.75 	& 2.06	  & 4.53	& 8.87	 \\
\multirow{4}{*}{}	& 3:4 &	1.752	&	1.957	& 2.29 	& 2.77    & 5.32	& 11.08	 \\
\hline
\multirow{4}{*}{Ti}	& 0:4 &	1.145	&	1.448	& - 	& -	      & -		& -		 \\
\multirow{4}{*}{}	& 1:4 &	0.930	&	0.997	& -0.84	&-1.78    & 0.09	& -0.12	 \\
\multirow{4}{*}{}	& 2:4 &	1.024	&	1.159	& -0.47	&-1.14    & 1.99	& 2.89	 \\
\multirow{4}{*}{}	& 3:4 &	1.196	&	1.511	& 0.20	&0.25	  & 3.09	& 5.68	 \\
\hline
\end{tabular}
\end{table}

\begin{figure}[h!]
  \centering
  \includegraphics[width=1.0\textwidth]{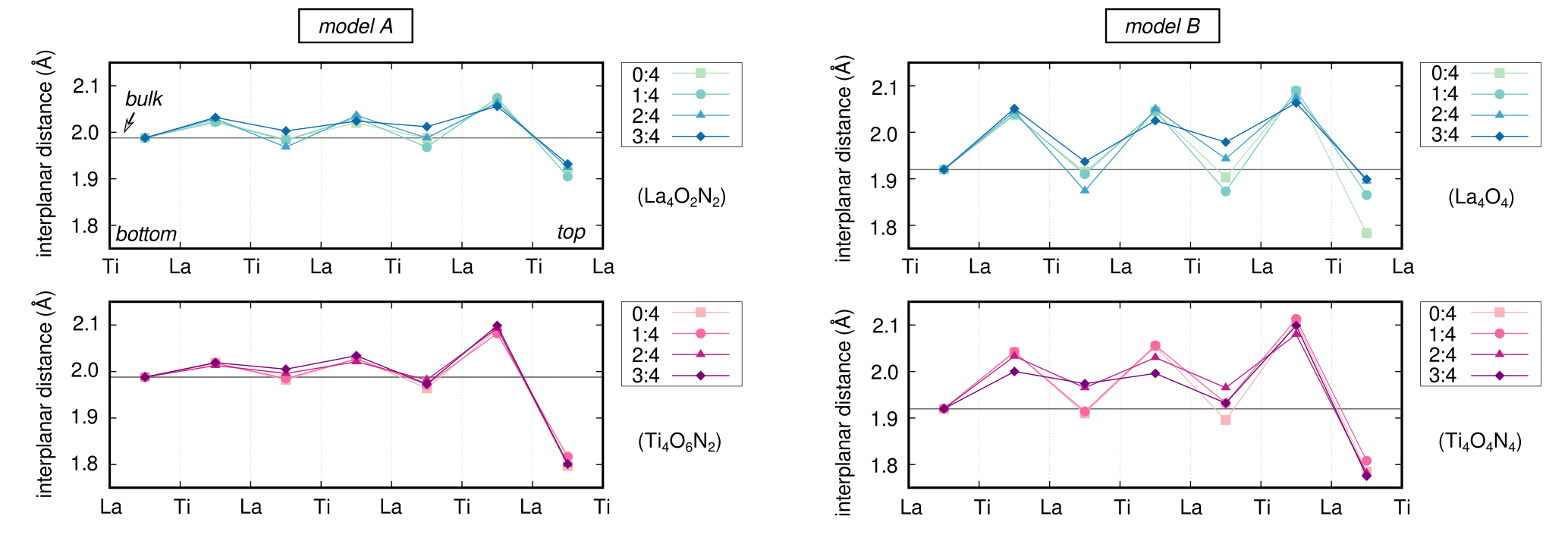}
  \caption{Displacements in the models A (left) and B (right) of the 2$\times$2$\times$4 simulation cell for the La (top panels) and Ti (bottom panels) terminations. The geometry optimization is carried out by relaxing all atoms, except for those in the bottommost unit cell, which are kept fixed at bulk positions.}
\label{fig:displacements}
\end{figure}

\begin{table}[h!]
\caption{Geometry characteristics of the surface layers in all slab models.}
\label{tbl:surf-geo}
\centering
\begin{tabular}{c|c|c|rr|cc}
\hline
Model & Termination & \textit{trans}:total & $\Delta _{1-2}$ (\%)$^a$ & $\Delta _{2-3}$ (\%)$^a$ & s$_O$ (\r{A})$^b$ & s$_N$ (\r{A})$^b$\\ 
\hline
\multirow{8}{*}{model A}& \multirow{4}{*}{La} 	& 0:4 & -3.2	& 3.9 	& 0.288			& 0.170		  \\
\multirow{8}{*}{}		 & \multirow{4}{*}{}	& 1:4 & -4.2	& 4.3	& -				& 0.105/0.119 \\
\multirow{8}{*}{}		 & \multirow{4}{*}{}	& 2:4 & -3.2	& 3.8	& -				& 0.100		  \\
\multirow{8}{*}{}		 & \multirow{4}{*}{}	& 3:4 & -2.8	& 3.4	& -				& 0.091/0.114 \\\cline{2-7}
\multirow{10}{*}{} 		 & \multirow{4}{*}{Ti} 	& 0:4 & -9.6	& 5.2	& -0.258/-0.181/0.332 & 0.075 \\
\multirow{8}{*}{}		 & \multirow{4}{*}{}	& 1:4 & -8.6	& 4.7	& -0.158/-0.058/0.128/0.371&- \\
\multirow{8}{*}{}		 & \multirow{4}{*}{}	& 2:4 & -9.4	& 5.3	& -0.103/0.009/0.072/0.314 &- \\
\multirow{8}{*}{}		 & \multirow{4}{*}{}	& 3:4 & -9.4	& 5.6	& -0.091/0.029/0.050/0.299 &- \\\cline{1-7}
\multirow{8}{*}{model B}& \multirow{4}{*}{La} 	& 0:4 & -7.1	& 4.6	& 0.387			& - 		  \\
\multirow{8}{*}{}		 & \multirow{4}{*}{}	& 1:4 & -2.8	& 4.8	& - 			& 0.132 	  \\
\multirow{8}{*}{}		 & \multirow{4}{*}{}	& 2:4 & -1.3	& 4.0	& - 			& 0.114 	  \\
\multirow{8}{*}{}		 & \multirow{4}{*}{}	& 3:4 & -1.1	& 3.4	& - 			& 0.105		  \\\cline{2-7}
\multirow{8}{*}{} 		 & \multirow{4}{*}{Ti} 	& 0:4 & -7.0	& 5.6	& -0.338		& 0.208 	  \\
\multirow{8}{*}{}		 & \multirow{4}{*}{}	& 1:4 & -5.8	& 5.9	& -0.170/0.324	& - 	 	  \\
\multirow{8}{*}{}		 & \multirow{4}{*}{}	& 2:4 & -7.2	& 4.2	& -0.146/0.306	& - 	 	  \\
\multirow{8}{*}{}		 & \multirow{4}{*}{}	& 3:4 & -7.5	& 5.2	& -0.142/0.303	& - 	 	  \\\cline{1-7}
\hline
\end{tabular}
\begin{flushleft}
$^a$ The interplanar distances, $\Delta$, are calculated as the difference between the averaged topmost atomic layer heights with respect to the bulk value along that specific direction.\\
$^b$ The rumpling, s, is calculated as the difference along z between the O or N atoms and the plane formed by the metal ones.\\
\end{flushleft}
\end{table}

\vspace{1.2cm}
In Figure \ref{fig:displacements}, we report the change of the interplanar distance as a function of the \textit{trans}:total ratio for both terminations and both models. As we can see, model A has generally smaller changes in the interplanar distances compared to model B. In general we observe an inward relaxation of the surface layer, followed by an oscillating inwards/outwards relaxation sequence. While the La termination of Model A shows a systematic change as a function of the \textit{trans}:total ratio going towards a more uniformly expanded surface without oscillations, the Ti termination shows no large changes. Model B surfaces in general show larger changes as a function of the \textit{trans}:total ratio also moving from oscillatory to more uniformly expanded with increasing number of \textit{trans} layers. The interplanar spacings between the first and second as well as second and third layer are numerically given in Table \ref{tbl:surf-geo}, along with the rumpling of oxygen and nitrogen atoms in the surface layer. In general, we see that the anion exchange decreases the rumpling in the surface layer.

\begin{figure}[h!]
  \centering
  \includegraphics[width=1.0\textwidth]{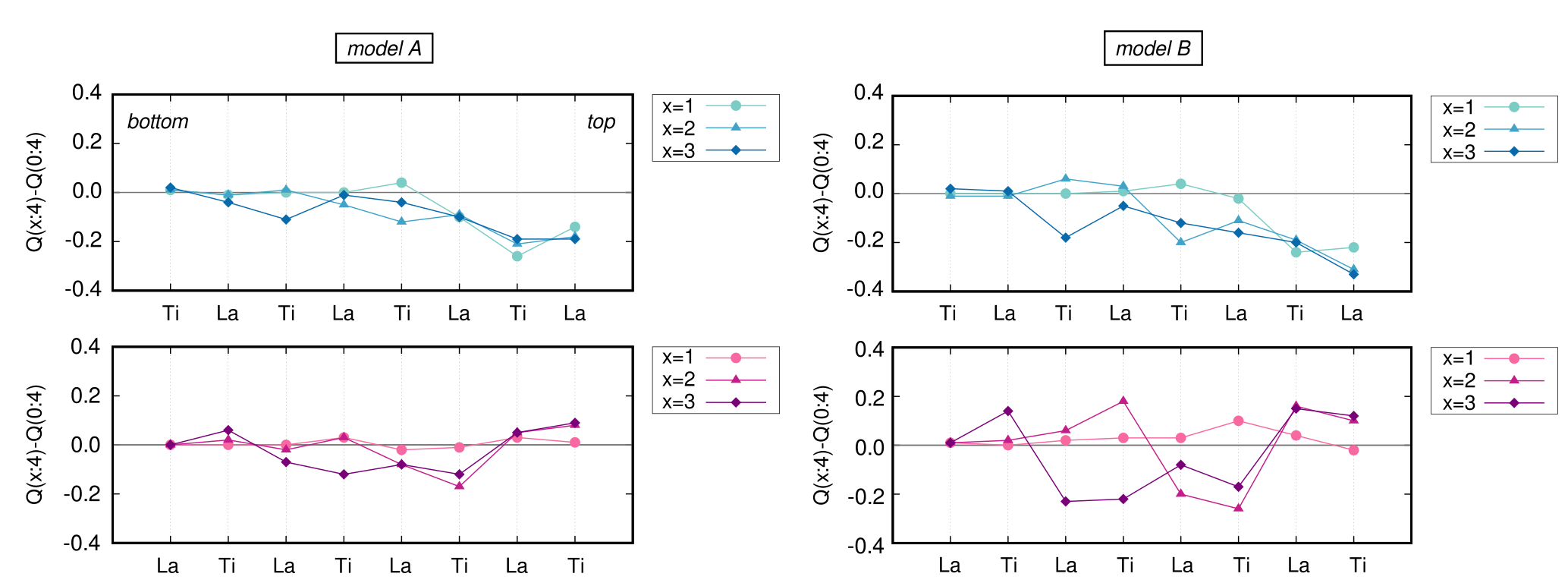}
  \caption{Change in the L\"{o}wdin charge per layer with the anion exchange for the two models in the 2$\times$2$\times$4 slab; La-terminated surfaces (top) and Ti-terminated ones (bottom).}
\label{fig:Lowdin_charges}
\end{figure}

In Figure \ref{fig:Lowdin_charges}, we report the change in layer-averaged L\"{o}wdin charges upon anion exchange. While for the La terminated surfaces of both model A and B, we see a large negative change (electron accumulation) in the exchanged surface region, that steadily decreases towards the bottom of the slab, the Ti termination shows a completely different behaviour. Starting from two \textit{trans} layers, we there see a positive change (hole accumulation) at the surface, followed by a negative region (electron accumulation) before becoming positive again. The lower boundary of this electron accumulation region shifts as a function of the number of exchanged layers. This is in line with the different electronic reconstruction observed for the two termination in the main text.

\begin{figure}[t]
  \centering
  \includegraphics[width=0.5\columnwidth]{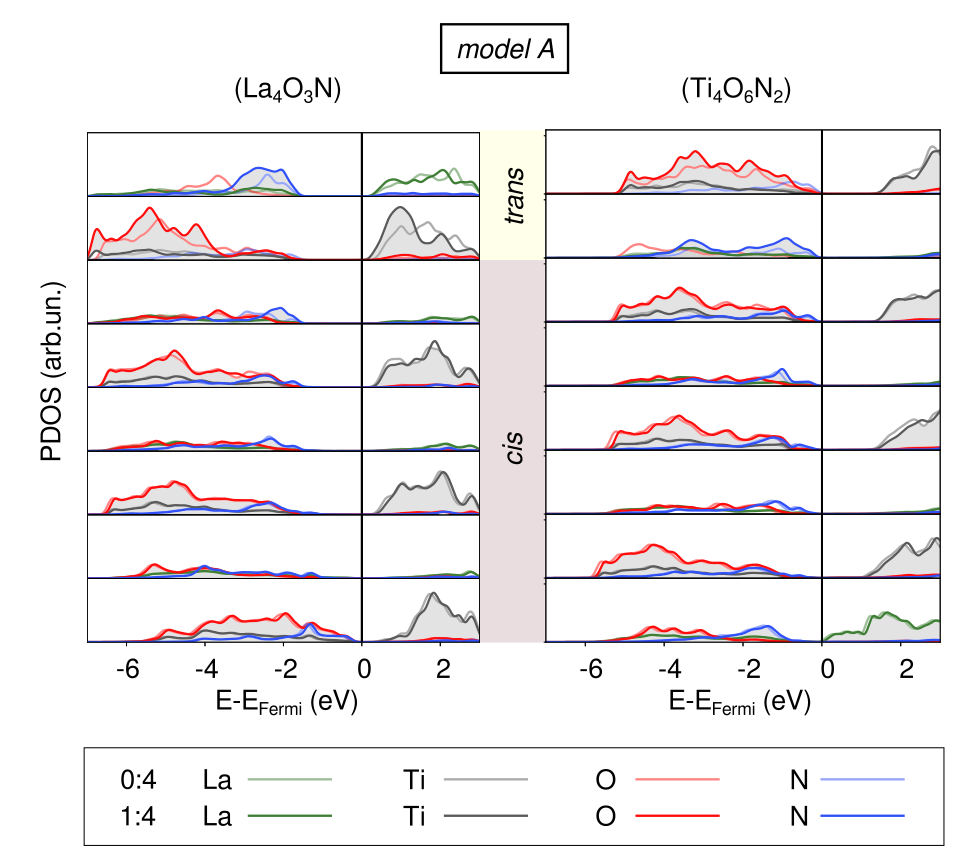}
  \caption{Layer-resolved PDOS of both terminations in model A, where the 1:4 slab is drawn together with the 0:4 slab. The PDOS of model B are in the main text, Figure \ref{fig:pdos-B}.}
\label{fig:pdos-A}
\end{figure}

In Figure \ref{fig:pdos-A} we report the layer-resolved projected density of states for model A. The changes as a result of the different anion order at the topmost surface layer are similar to those observed in model B (see Figure \ref{fig:pdos-B} in the main text), however, pronounced to a lesser extent.

\clearpage
%%%%%%%%%%%%%%%%%%%%%%%%%%%%%%%%%%%%%%%%%%%%%%%%%%%%%%%%%%%%%%%%%%%
\subsection{Additional surface models}
%%%%%%%%%%%%%%%%%%%%%%%%%%%%%%%%%%%%%%%%%%%%%%%%%%%%%%%%%%%%%%%%%%%

\begin{figure}[h]
  \centering
  \includegraphics[width=1.0\textwidth]{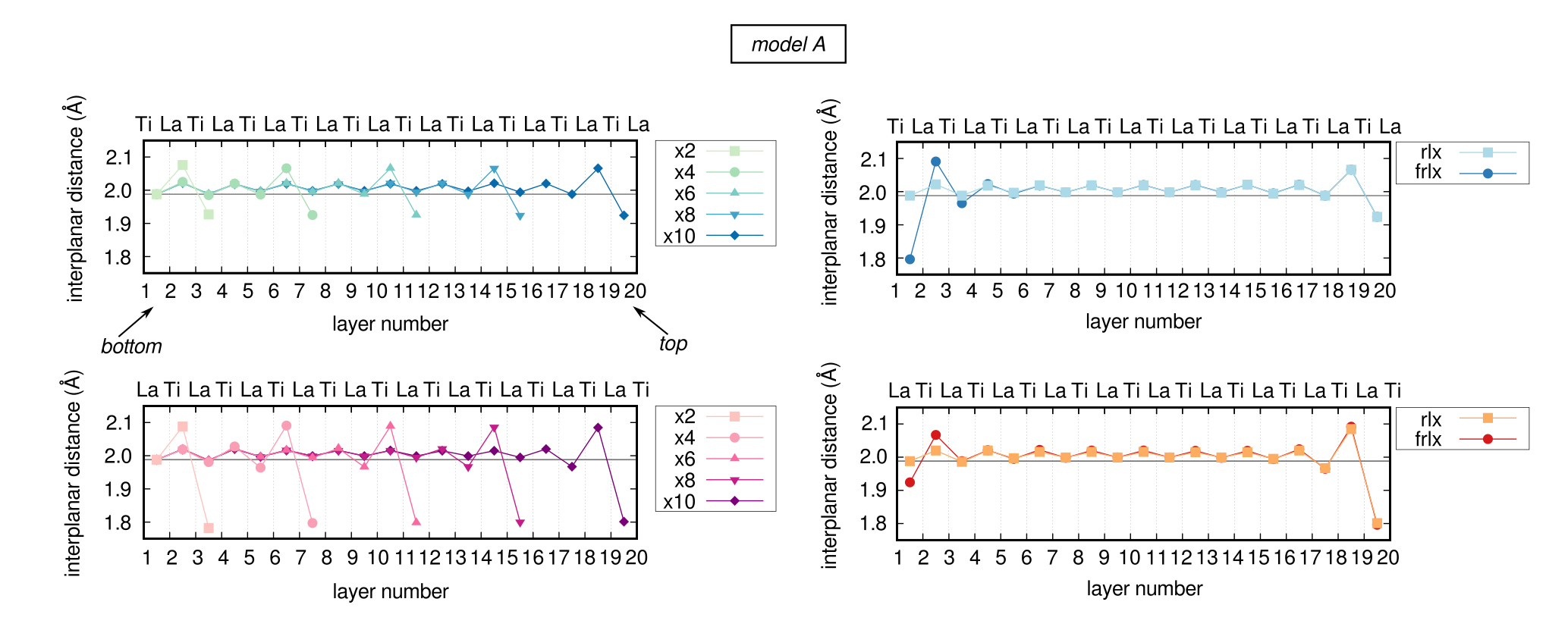}
  \caption{Left column: interplanar spacings in the lateral $\sqrt{2}/2\times\sqrt{2}/2$-R45$^{\circ}$ simulation cell with $n$ unit-cell layers thickness of the slab model. All atoms were relaxed, except for those in the bottommost unit cell, which were kept frozen at their bulk positions. Right column: comparison of interplanar spacings between a fully relaxed 10-layers slab (frlx, no frozen atoms) and one where the bottommost unit cell is kept fixed as in our standard setup (rlx). Top row: La-termination, bottom row: Ti-termination.}
\label{fig:displacements-sq}
\end{figure}

Here we present surface calculations obtained with slab models that have dimension $\sqrt{2}/2\times\sqrt{2}/2$-R45$^{\circ}$ with respect to the $2\times2$ supercell used in the above calculations. The reduced lateral dimensions allows us to consider thicker slab models to verify if our main conclusions are affected by artefacts resulting from our slab setup (4x model used in the main text). As shown for Model A La- and Ti-terminated surfaces in the left column of Figure \ref{fig:displacements-sq}, the relaxation of the surface layer is not affected by the limited thickness of our slab as we observe an upwards shift of the same relaxation pattern at the surface and the underlying layers with increasing slab thickness. As shown in the right column of \ref{fig:displacements-sq}, keeping the bottom layer frozen does also not affect the relaxation pattern at the surface.

In Table \ref{tbl:exch-en-all}, we show the anion exchange energies without ($E_{ionex}^{urlx}$) and with ($E_{ionex}$) relaxation for both Model A and B as well as the La and Ti termination. We consider slabs that have the asymmetric geometry with a fixed bottom layer as in our standard setup (asymm), an asymmetric slab, where the fixing of the bottom layer is omitted (frlxasymm) and a symmetric slab setup with one additional atomic layer at the bottom of the slab (symm). We can see that while for model A there is a large difference between the symmetric and asymmetric slabs without relaxation, this difference is reduced to almost nothing when the models are relaxed. We can thus see that the different slab setups lead to very similar $E_{ionex}$ for the different models and terminations, slight differences occurring for 3:4 ratios when the bottom of the slab is allowed to relax. More importantly though the slab setup does not affect our qualitative conclusions that only the Ti terminated surface has negative $E_{ionex}$ for up to two layers.

\begin{table}[h]
\caption{Exchange energies for the two models A: La$_4$O$_2$N$_2$/Ti$_4$O$_6$N$_2$ and B: La$_4$O$_4$/Ti$_4$O$_4$N$_4$. The $E_{ionex}^{urlx}$ determines the energy difference between the relaxed 0:4 models and the unrelaxed mixed one, where just the anion exchange has occurred. E$_{ionex}$, on the other hand considers the relaxation as well.}
\label{tbl:exch-en-all}
\centering
\begin{tabular}{c|c|cc|c|ccc|cc}
\hline
 & & \multicolumn{3}{c|}{E$_{ionex}^{urlx}$\,(eV)} & \multicolumn{5}{c}{E$_{ionex}$\,(eV)}\\
\hline
 &\textit{trans}:total& \multicolumn{2}{c|}{model A} &\multicolumn{1}{c|}{model B} & \multicolumn{3}{c|}{model A} & \multicolumn{2}{c}{model B} \\
\hline
Termination     	& (layers) & asymm & symm & asymm & asymm & frlxasymm & symm  & asymm & frlxasymm \\
\hline
\multirow{3}{*}{La}	& 1:4  & 2.25  & 2.70 & 5.40  & 1.15  & 1.15   & 1.01  & 1.32  & 1.48 \\
\multirow{3}{*}{}	& 2:4  & 4.53  & 4.51 & 8.87  & 1.75  & 1.76   & 1.57  & 2.06  & 2.30 \\
\multirow{3}{*}{}	& 3:4  & 5.32  & 6.44 & 11.08 & 2.29  & 2.11   & 2.12  & 2.77  & 2.59 \\
\hline
\multirow{3}{*}{Ti}	& 1:4  & 0.09  & 1.84 & -0.12 & -0.84 & -0.83  & -0.82 & -1.78 & -1.86 \\
\multirow{3}{*}{}	& 2:4  & 1.99  & 3.48 & 2.89  & -0.47 & -0.52  & -0.39 & -1.14 & -1.29 \\
\multirow{3}{*}{}	& 3:4  & 3.09  & 5.38 & 5.68  & 0.20  & 0.12   & 0.24  & 0.25  & -1.08 \\
\hline
\end{tabular}
\end{table}

\newpage
In order to evaluate if the proximity of the frozen bottom layer or the small thickness of the slab could have an effect on the observed $E_{ionex}$, we repeat the calculations for model A also with a thicker 6-layer slab. The results shown in Table \ref{tbl:surf-en6}, show that the results are qualitatively and even quantitatively the same, only the Ti termination having negative $E_{ionex}$ for up to two \textit{trans} layers at the surface. We thus conclude that the slab setup used in the main text has no effect on our conclusions as the same results are obtained with thicker or differently constructed slabs.

\begin{table}[h]
\caption{Surface and anion exchange energies of the 6-layered model A slab with different \textit{trans}:total ratios. The same lattice parameters are used as for the 4-layered slab.}
\label{tbl:surf-en6}
\centering
\begin{tabular}{c|c|cc}
\hline
\multirow{2}{*}{Termination} & Mixing ratio & \multicolumn{2}{c}{model A} \\
\multirow{2}{*}{} &  Mixing ratio & $E_{surf}$\,(J/m$^2$)$^a$ & E$_{ionex}$\,(eV)\\
\hline
\multirow{6}{*}{La}	& 0:6	&	1.115	&	-	    \\
\multirow{6}{*}{}	& 1:6	&	1.408	&	1.14	\\
\multirow{6}{*}{}	& 2:6	&	1.570	&	1.77	\\
\multirow{6}{*}{}	& 3:6	&	1.708	&	2.31	\\
\multirow{6}{*}{}	& 4:6	&	1.847	&	2.84	\\
\multirow{6}{*}{}	& 5:6	&	1.974	&	3.34	\\
\hline
\multirow{6}{*}{Ti}	& 0:6	&	1.104	&	-	    \\
\multirow{6}{*}{}	& 1:6	&	0.891	&	-0.82	\\
\multirow{6}{*}{}	& 2:6	&	0.983	&	-0.47	\\
\multirow{6}{*}{}	& 3:6	&	1.123	&	0.07	\\
\multirow{6}{*}{}	& 4:6	&	1.267	&	0.64	\\
\multirow{6}{*}{}	& 5:6	&	1.436	&	1.29	\\
\hline
\end{tabular}
\end{table}

%%%%%%%%%%%%%%%%%%%%%%%%%%%%%%%%%%%%%%%%%%%%%%%%%%%%%%%%%%%%%%%%%%%
\section{Polar interface} \label{sec:polar-interface}
%%%%%%%%%%%%%%%%%%%%%%%%%%%%%%%%%%%%%%%%%%%%%%%%%%%%%%%%%%%%%%%%%%%

\begin{figure}[h]
  \centering
  \includegraphics[width=1.0\textwidth]{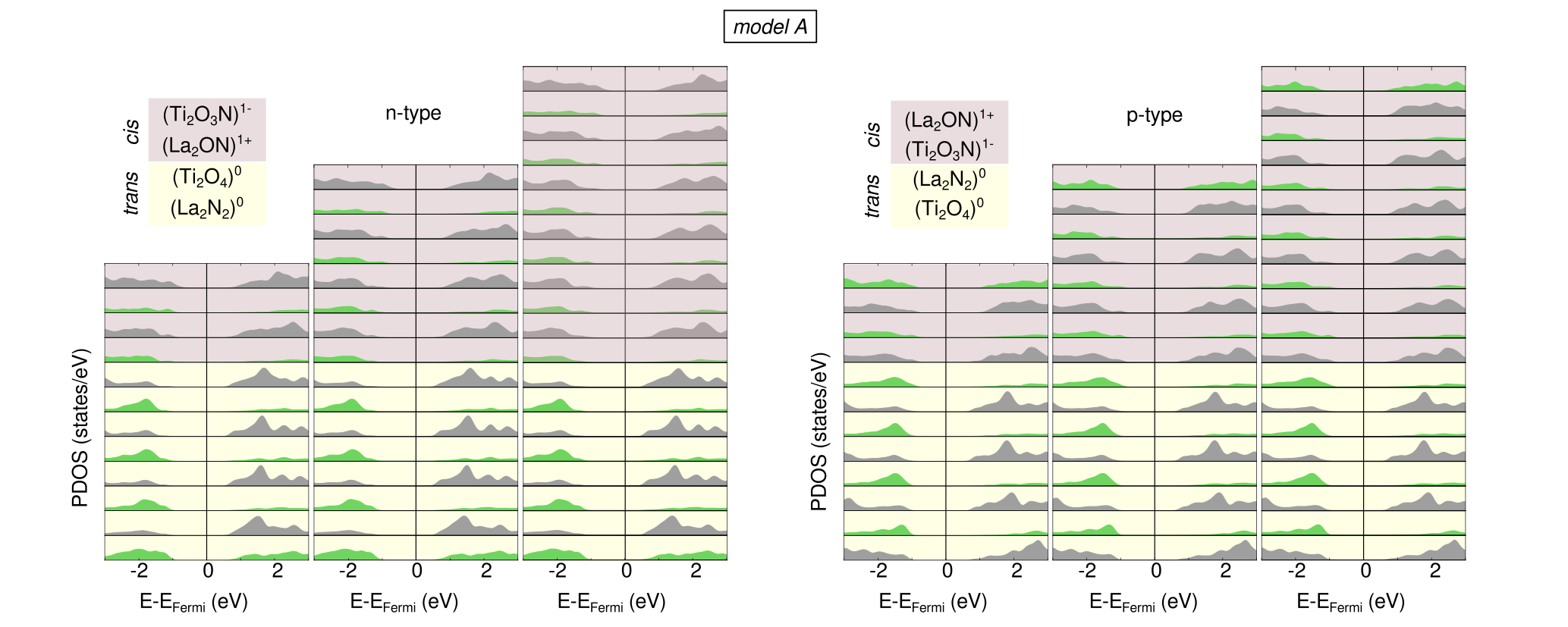}
  \caption{Layered total PDOS of the polar interface for model A, where for each termination different thickness of the \textit{cis}-ordered slab was tested. In green we indicate the La-atomic layers and in grey - the Ti ones. The p-type interface is on the left and the n-type on the right. The corresponding data for Model B is presented in the main text, Figure \ref{fig:pdos-B-laosto}.}
\label{fig:pdos-A-laosto}
\end{figure}

In Figure \ref{fig:pdos-A-laosto} we show the layer-resolved projected density of states of the polar interfaces constructed from model A bulk. The polarity in this model is weaker and it fails to reach band-inversion and electron/hole doping of the interface for the overlayer thicknesses investigated here.

\begin{figure}[h]
  \centering
  \includegraphics[width=1.0\textwidth]{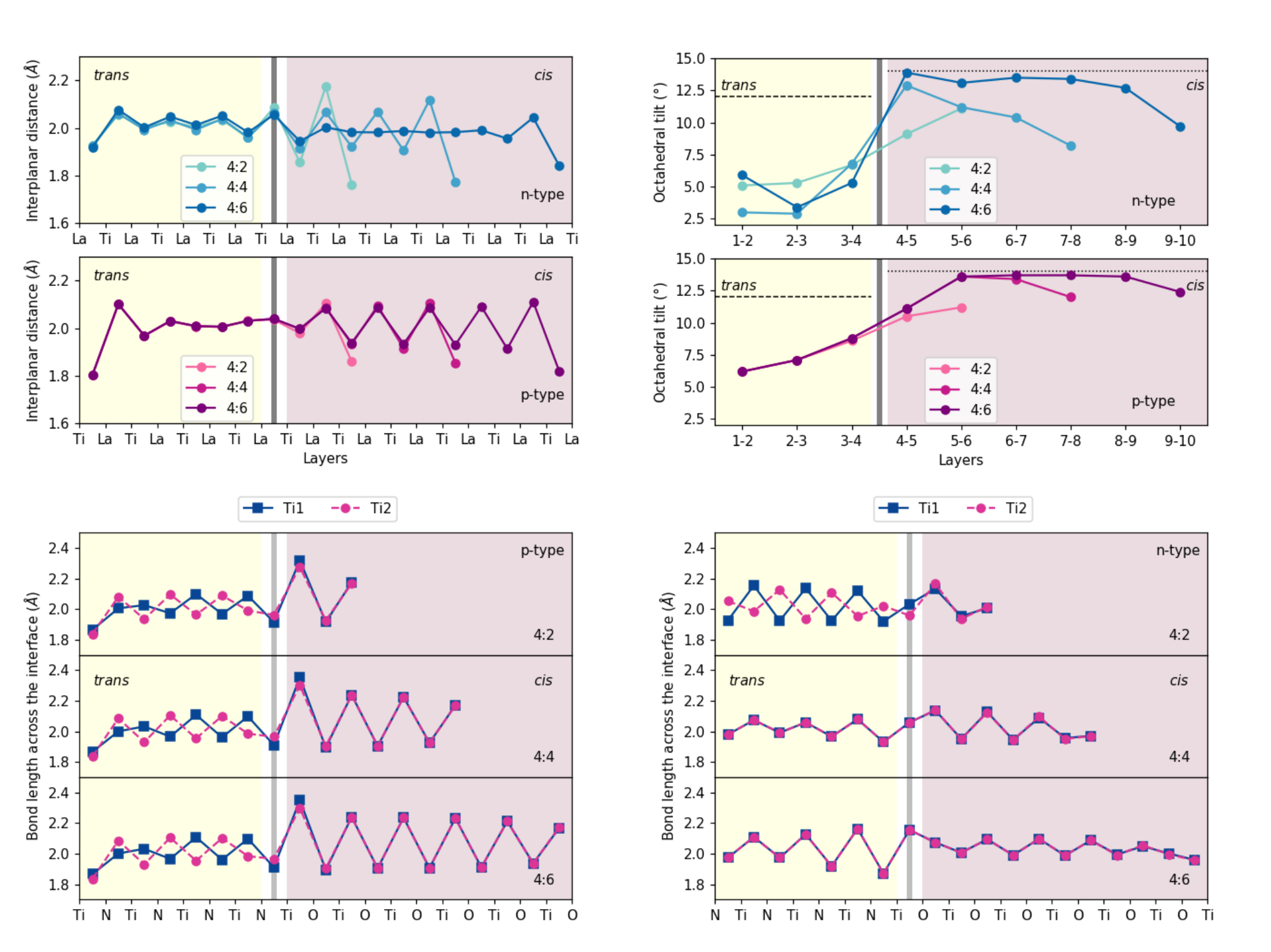}
  \caption{Interplanar distances, octahedral tilts and Ti-X (X=O,N) bond lengths along the normal of the interface in Model B  p-type and n-type interfaces. Ti1 and Ti2 correspond to each of the two Titanium atoms within each layer of the polar-interface models.}
\label{fig:geometry-laosto}
\end{figure}

In Figure \ref{fig:geometry-laosto}, we analyze the geometry of the model B interfaces in terms of interlayer spacings,  octahedral tilts and Ti-X bond lengths along the interface normal direction. In terms of inter-layer spacings, both the n- and p-type interface show strong oscillations in the polar \textit{cis} part of the slab for thin overlayers. Interestingly though, these oscillations completely disappear once the slab becomes solidly metallic for the 4:6 n-type interface. We see in general strongly suppressed octahedral rotations in the \textit{trans} part of the interface compared to the corresponding bulk value. For the n-type interface the bulk tilt angle is only recovered for the metallic 4:6 slab, whereas for the p-type interface a bulk-like structure is already recovered for the 4:4 slab. Besides changes in octahedral rotation magnitudes, we also observe a favoring of in-phase octahedral rotations around the interface normal direction in the \textit{trans} part of the slab, which is complete for the n-type and partial for the p-type interface. 

Similarly to other oxide interfaces, the strong suppression of octahedral tilts in the non-polar part of the interface can lead to the appearance of polar distortions to optimise bonding in the perovskite structure.\cite{Gazquez2017} We see these polar distortions emerging as strong oscillations of the Ti-X bond lengths, where up- and down-oriented bonds of the same Ti differ by about 0.24\AA. Such polarity was also seen in the \textit{trans}-ordered bulk.\cite{Vonrueti2018b} For all p-type interfaces as well as the 4:2 n-type interface we observe opposite displacements of Ti-X bonds for different Ti in the same layer. This implies an anti-ferroelectric type distortion, which will however have a similar effect on orbital energies as a purely ferroelectric one observed in n-type 4:4 and 4:6.

\bibliographySI{library.bib}

\end{document}